\documentclass[twocolumn,secnumarabic,amssymb, nobibnotes, aps, prl, superscriptaddress]{revtex4-2}

\usepackage[english]{babel}

\newcommand*\vtick{\textsc{\char13}}
\usepackage[]{graphicx}
\usepackage{tabularx}
\usepackage[usenames,dvipsnames]{color}
\usepackage{soul}
\usepackage[normalem]{ulem} 
\usepackage{bm}
\usepackage{amsmath}
\usepackage{lineno}
\usepackage{gensymb}
\usepackage{tabularx}
\usepackage{amsfonts}
\usepackage{mathrsfs}
\usepackage{graphicx}
\usepackage{dcolumn}
\usepackage{bm}
\usepackage{color}
\usepackage[colorlinks,bookmarks=false,citecolor=blue,linkcolor=red,urlcolor=blue]{hyperref}

\usepackage{multirow}
\usepackage{physics}
\usepackage{multirow}
\begin{document}

\title{Signatures of time-reversal-symmetry breaking in multiband 2H-TaS$_2$ revealed by zero-field Josephson nonreciprocity}

\author{Daniel Margineda}
\affiliation{CIC nanoGUNE BRTA, Donostia-San Sebastián, Spain}

\author{David Caldevilla-Asenjo}
\affiliation{Centro de Física de Materiales – Materials Physics Center (CFM-MPC) CSIC-EHU, San Sebastian, 20018, Spain}

\author{Yuriy Yerin}
\affiliation{Centro de Física de Materiales – Materials Physics Center (CFM-MPC) CSIC-EHU, San Sebastian, 20018, Spain}
\affiliation{Istituto di Struttura della Materia, Consiglio Nazionale delle Ricerche, via Salaria Km~29.3, I-00016 Monterotondo Stazione, Italy}

\author{Covadonga  \'Alvarez-Garc\'ia}
\affiliation{CIC nanoGUNE BRTA, Donostia-San Sebastián, Spain}

\author{Andrei Mazanik}
\affiliation{Centro de Física de Materiales – Materials Physics Center (CFM-MPC) CSIC-EHU, San Sebastian, 20018, Spain}

\author{Maxim Ilyn}
\affiliation{Centro de Física de Materiales – Materials Physics Center (CFM-MPC) CSIC-EHU, San Sebastian, 20018, Spain}
\author{Celia Rogero}
\affiliation{Centro de Física de Materiales – Materials Physics Center (CFM-MPC) CSIC-EHU, San Sebastian, 20018, Spain}

\author{Luis E. Hueso}
\affiliation{CIC nanoGUNE BRTA, Donostia-San Sebastián, Spain}

\author{F. Sebastian Bergeret}
\affiliation{Centro de Física de Materiales – Materials Physics Center (CFM-MPC) CSIC-EHU, San Sebastian, 20018, Spain}

\author{Marco Gobbi}
\affiliation{Centro de Física de Materiales – Materials Physics Center (CFM-MPC) CSIC-EHU, San Sebastian, 20018, Spain}

\begin{abstract}
Superconductors that spontaneously break time-reversal symmetry host complex order parameters and are widely regarded as a hallmark of unconventional superconductivity. Whether such symmetry breaking can also arise in superconductors with nominally isotropic spin-singlet pairing remains an open question. Here we report a zero-field Josephson diode effect in noncentrosymmetric 2H-TaS$_2$/2H-NbSe$_2$ van der Waals junctions. The diode efficiency shows no systematic correlation with supercurrent amplitude, TaS$_2$ thickness, or normal-state resistance, arguing against simple extrinsic, purely interfacial, or transparency-driven mechanisms. Time-reversal-symmetric scenarios are further tested using symmetry-controlled and molecule-intercalated control devices, in which the nonreciprocal response is absent or strongly reduced. Normal-state Hall transport in TaS$_2$ exhibits a nonlinear response consistent with multiband correlated electronic states. Within a Josephson framework, our modelling shows that interband scattering acts as a phase-locking mechanism generating an intrinsic anomalous phase difference and a nonsinusoidal asymmetric current-phase relation, leading to finite zero-field rectification. Together, zero-field Josephson nonreciprocity and nonlinear Hall transport provide complementary evidence for a multiband superconducting phase structure in 2H-TaS$_2$, consistent with intrinsic time-reversal-symmetry breaking.

\end{abstract}
\keywords{van der Waals superconductors, Ising superconductivity, organic intercalation, Time-reversal symmetry breaking, Multiband superconductivity, Josephson diode effect, Nonreciprocal transport, S+is' pairing state}

\maketitle

\begin{figure*}[ht]
\centering
\includegraphics[scale=0.28 ]{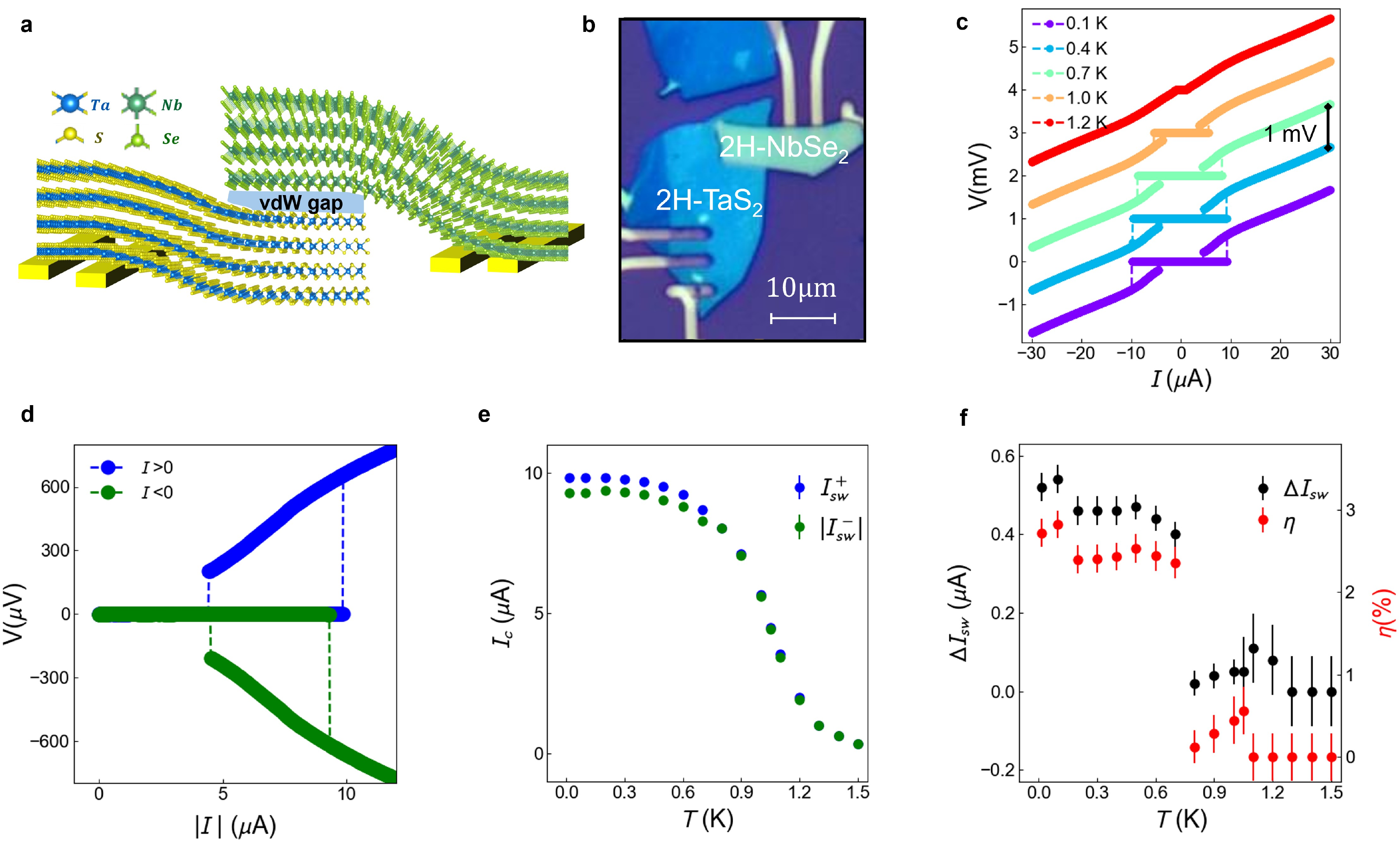}
\caption{
\textbf{Zero-field Josephson diode effect in van der Waals heterojunction.} 
\textbf{a,} Schematic of the Josephson junction under study, consisting of 2H-TaS$_2$ and 2H-NbSe$_2$ superconducting electrodes with the van der Waals gap operating as a weak link. \textbf{b,} Optical micrograph of the Josephson junction. Bottom Au contacts allow the characterization of the interface.
\textbf{c,} Current-voltage characteristics for representative temperatures showing hysteretic behavior of underdamped Josephson junctions.
\textbf{d,} Zero-field Josephson diode effect. Current–voltage characteristics taken at $T=100\, \text{mK}$ show different switching currents depending on the positive (blue dots) and negative (green dots) bias current. \textbf{e,} Temperature dependence of the positive $I_{sw}^+$ and negative $|I_{sw}^-|$ switching currents. Nonreciprocal transport $\Delta I_{sw}=I_{sw}^+-|I_{sw}^-| \neq 0$ persists up to approximately $T=0.6T_c$.  \textbf{f,}  Temperature dependence of $\Delta I_{sw}$ (black dots) and the Josephson diode efficiency $\eta$ (red dots). Error bars are taken from the Full Width at Half Maximum (FWHM) of the switching current statistics taken at specific temperatures.} 
\label{fig1} 
\end{figure*}

 Symmetries play a central role in defining superconducting states. In most superconductors, phonon-mediated pairing yields a real order parameter that preserves inversion and time-reversal symmetry (TRS). The spontaneous breaking of TRS is therefore exceptional and signals the emergence of unconventional superconducting states, often associated with complex or multicomponent order parameters arising from non-phononic pairing mechanisms. Identifying robust experimental signatures of TRS breaking remains a central challenge in condensed-matter physics.
 
On the other hand, the superconducting diode effect (SDE), defined as the appearance of nonreciprocal critical currents in Josephson junctions or bulk superconductors, has emerged as a powerful symmetry-sensitive probe~\cite{nad23,ma25}. Unlike its dissipative counterpart, superconducting nonreciprocity is understood to require the simultaneous breaking of inversion symmetry and TRS to produce unequal critical currents depending on the bias direction ($I_c^+\neq I_c^-$)~\cite{yua22,dav22,zha22,he22,pal22}.
First observed in noncentrosymmetric superconductors~\cite{wak17,and20,zha20}, the SDE has proven instrumental in identifying mechanisms that lift inversion symmetry under an applied magnetic field, including strong spin-orbit coupling~\cite{bau22,mar23,baur22,Costa2023,lot24}, geometric asymmetries,~\cite {baur22,lyu21A,hou23,cas25} vortex-induced effects~\cite{ gol22,mar23,Fukaya2025,gut23} and nonequilibrium conditions~\cite{mar23a,sha25}. By contrast, the zero-field superconducting diode effect (ZF-SDE) is interpreted as a signature of spontaneous TRS breaking, as reported in magnetically proximitized superconductors~\cite{Narita2022,tra23,jeo22,li25}, valley-polarized states in twisted graphene~\cite{lin22,mer23,Hu2023} and materials proposed to host unconventional pairing symmetries~\cite{Qiu2023,Le2024,zhao23,Qi2025,Anwar2023}. In these systems, zero-field nonreciprocity originates from superconducting states with an internal phase structure capable of generating directional supercurrents.
However, nonreciprocal superconducting transport can also be realized in engineered Josephson circuits, such as multiterminal or back-action devices, where rectification arises from device architecture and effective current-phase-relation engineering rather than from intrinsic time-reversal-symmetry breaking of the superconducting condensate~\cite{mar25,gup23,lup24}. Moreover, recent observations of spontaneous nonreciprocal transport in transition-metal dichalcogenide (TMD) heterostructures~\cite {Wu2022,24_chiral,liu24,Ma2025,chiral4Hb}, which host isotropic s-wave spin-singlet gaps~\cite{sim24}, have questioned whether TRS breaking is strictly required. Alternative mechanisms, including carrier-type asymmetries~\cite {Wu2022} and strain-induced effects~\cite{liu24,Ma2025} have been proposed. A central challenge is therefore to identify experimental observables that distinguish multiband superconducting states preserving time-reversal symmetry from those in which time-reversal symmetry is spontaneously broken, particularly in TMDs with nominally conventional pairing but complex electronic structure.

Here, we address this question in 2H-TaS$_2$/2H-NbSe$_2$ van der Waals Josephson junctions. A robust zero-field Josephson diode is systematically probed via switching-current statistics across multiple devices. 
The nonreciprocal signal remains finite and reproducible irrespective of TaS$_2$ thickness, excluding interface-dominated or strain-induced mechanisms. 

Control experiments further clarify the underlying symmetries: homojunctions composed of the same superconducting material exhibit reciprocal transport, indicating that broken inversion symmetry is necessary but not sufficient for a measurable Josephson diode effect, even in twisted structures with microscopically broken inversion symmetry.
Hybrid Josephson junctions with engineered carrier-type asymmetry are realized through selective molecular intercalation of TaS$_2$; the resulting reduction of diode efficiency is inconsistent with an interfacial charge depletion mechanism~\cite {Hu2007,mis21}. Instead, the combined observation of zero-field Josephson nonreciprocity and nonlinear Hall transport in 2H-TaS$_2$ points to the coexistence of interacting electron- and hole-derived condensates consistent with a multiband superconducting state with an intrinsic phase structure. We support this interpretation with a theoretical model in which interband scattering acts as a phase-locking channel, generating mixed-gradient (drag) terms and an anomalous phase difference that stabilizes a complex $s+is'$ state. Within this framework, we identify the interband scattering amplitude as the key control parameter governing the nonreciprocal response in nominally conventional multiband superconductors.

\begin{figure*}[ht]
\centering
\includegraphics[scale=0.26 ]{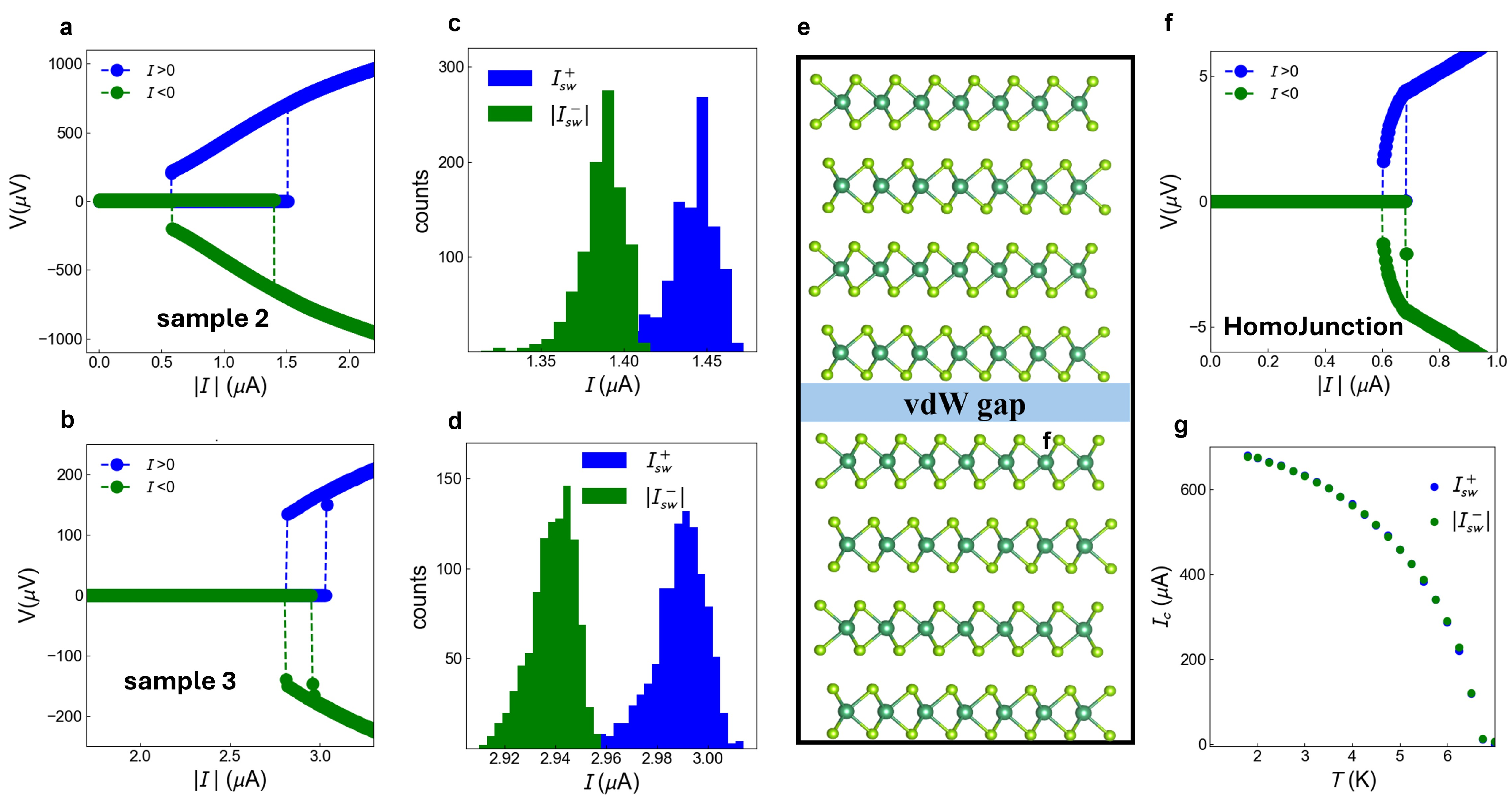}
\caption{
\textbf{Zero-field Josephson diode reproducibility.} Nonreciprocal transport for sample 2 (\textbf{a,}) and sample 3 \textbf{b,} measured at $T=100\, \text{mK}$. \textbf{c,d} Switching current statistics for sample 2 and sample 3 respectively. \textbf{e,} Control experiment in inversion-symmetric NbSe$_2$-NbSe$_2$ junctions. \textbf{f,} IV characteristics of the control sample showing no finite $\eta$ within experimental resolution. \textbf{g,} Temperature dependence of the positive and negative switching current of the control sample. The absence of nonreciprocity in symmetric junctions establishes the requirement of broken inversion symmetry at the heterostructure level and excludes trivial geometric or measurement artefacts.}
\label{fig2} 
\end{figure*}
We fabricated van der Waals Josephson junctions composed of vertically stacked metallic 2H-TaS$_2$ and 2H-NbSe$_2$ crystals. The top NbSe$_2$ flake, with thickness $t\sim 60-80$ nm, is chosen to ensure a noncentrosymmetric heterostructures with broken inversion symmetry, while bottom Au contacts enable vertical transport across the junction (Fig.~\ref{fig1}a,b).
Upon sweeping the bias current, the device switches from the dissipationless superconducting state ($V=0$) to the normal state at a critical current $I_c$, and returns to the superconducting state at a lower retrapping current $I_r$ upon decreasing the bias current. The resulting hysteresis (Fig.~\ref{fig1}c) confirms 
Josephson coupling between the superconductors across the van der Waals gap. We extract a Stewart-McCumber parameter $\beta_c\approx (4I_c/\pi I_r)^2 =5$, consistent with an underdamped junction with capacitance $C=0.11\, \text{pF}$. A clear nonreciprocity in the switching current is observed: the switching currents for opposite bias directions differ, $I_{sw}^+ \neq|I_{sw}^-|$ (Fig.~\ref{fig1}d), constituting the hallmark of a zero-field superconducting diode effect. The nonreciprocal signal $\Delta I_{sw}=I_{sw}^+ - |I_{sw}^-|$ remains sizeable up to $0.7\, \text{K}$ ($\sim0.6 T_c$) corresponding to a diode efficiency $\eta=(I_{sw}^+- |I_{sw}^-|)/(I_{sw}^++ |I_{sw}^-|)\approx 3\%$.
To assess reproducibility and parameter dependence, we fabricated multiple devices with varying TaS$_2$ thickness. Josephson coupling and zero-field Josephson diode effect are observed in all samples, demonstrating the robustness of the effect. In underdamped Josephson junctions, the difference $I_c> I_r$ arises from local heating during switching~\cite{cuo08}. In this regime, stochastic phase slips broaden the distribution of switching currents $I_{sw}$ and could potentially mimic nonreciprocity. To exclude this possibility, we measured switching-current probability distributions (SCPD) at different temperatures for all devices. Figure~\ref{fig2}a-d shows representative current-voltage characteristics and SCPDs at $T=100\, \text{mK}$. The diode signal is determined from the difference between the modes of the distributions $I_c^{\pm}=\hat{I}_{sw}^{\pm}$. With increasing temperature, the distributions broaden due to the crossover from macroscopic quantum tunnelling to thermally activated phase-slips processes~\cite{pug20} (Extended Data Fig.\,1). Importantly, the nonreciprocal shift $\Delta_{sw}$ remains well separated from the distribution width for $T \leq 0.6 T_c$, ruling out stochastic switching as its origin. At the same time, the separation between the distribution modes decreases with temperature, evidencing a progressive suppression of the nonreciprocal response.

Control experiments were performed on homojunctions. In NbSe$_2$/NbSe$_2$ devices (Fig.~\ref{fig2}e), the switching currents remain reciprocal over the entire temperature range  (Fig.~\ref{fig2}f,g), demonstrating that broken inversion symmetry is necessary but not sufficient for a measurable Josephson diode effect. Similar results are obtained in TaS$_2$/TaS$_2$ homojunctions (Extended Data Fig.\,2). Even in twisted homojunctions away from high-symmetry angles, where inversion symmetry may be microscopically broken, a finite diode response is not guaranteed.

\begin{figure*}[ht]
\centering
\includegraphics[scale=0.26 ]{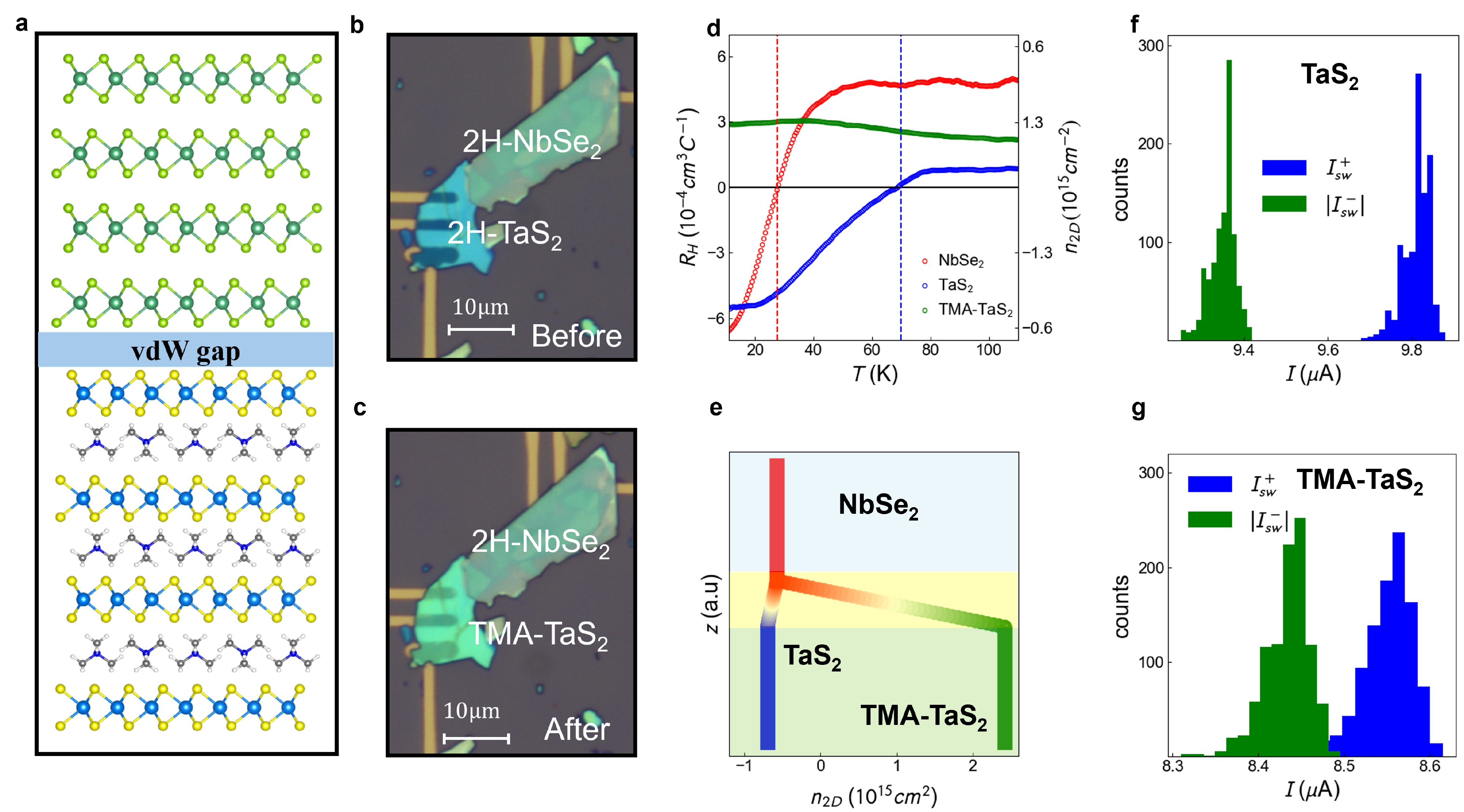}
\caption{
\textbf{Zero-field superconducting diode in molecule-TaS$_2$ Josephson junction.} \textbf{a,} Schematic of the hybrid Josephson junction consisting in TMA-intercalated TaS$_2$ and NbSe$_2$ top electrode.   Optical micrograph of the vdW Josephson junction before \textbf{b,} and after \textbf{c,} molecular intercalation. Color contrast is highly sensitive to thickness change upon intercalation from d$_p$=0.6\,nm to d$_{int} \simeq$ 1.0\,nm. \textbf{d,} Temperature dependence of the Hall coefficient for NbSe$_2$ (red dots), TaS$_2$ (blue dots) and TMA-TaS$_2$. Sign change in R$_H$ corresponds to carrier-type transition from hole-to-electron dominated transport below T$_{CDW}$ (dashed lines). For TMA-TaS$_2$, monolayer-like suppression of CDW order leads to electronic band reconfiguration and hole transport across the full temperature range (green dots). \textbf{e,} Representation of the charge distribution across the hybrid Josephson junction before and after intercalation (color scheme is applied as in panel \textbf{d,}). Upon intercalation, charge depletion region transitions from a p-type  (green line) to a n-type (red color) superconductor. Switching current statistics before  \textbf{f,} and after intercalation  \textbf{g,}. Nonreciprocal transport given by $\Delta_{sw}=I_sw^+ - I_{sw} \neq 0$ notably decreases upon intercalation, ruling out charge polarisation as the origin of asymmetric Josephson tunnelling.}
\label{fig3} 
\end{figure*}
\begin{figure*}[ht]
\centering
\includegraphics[scale=0.27 ]{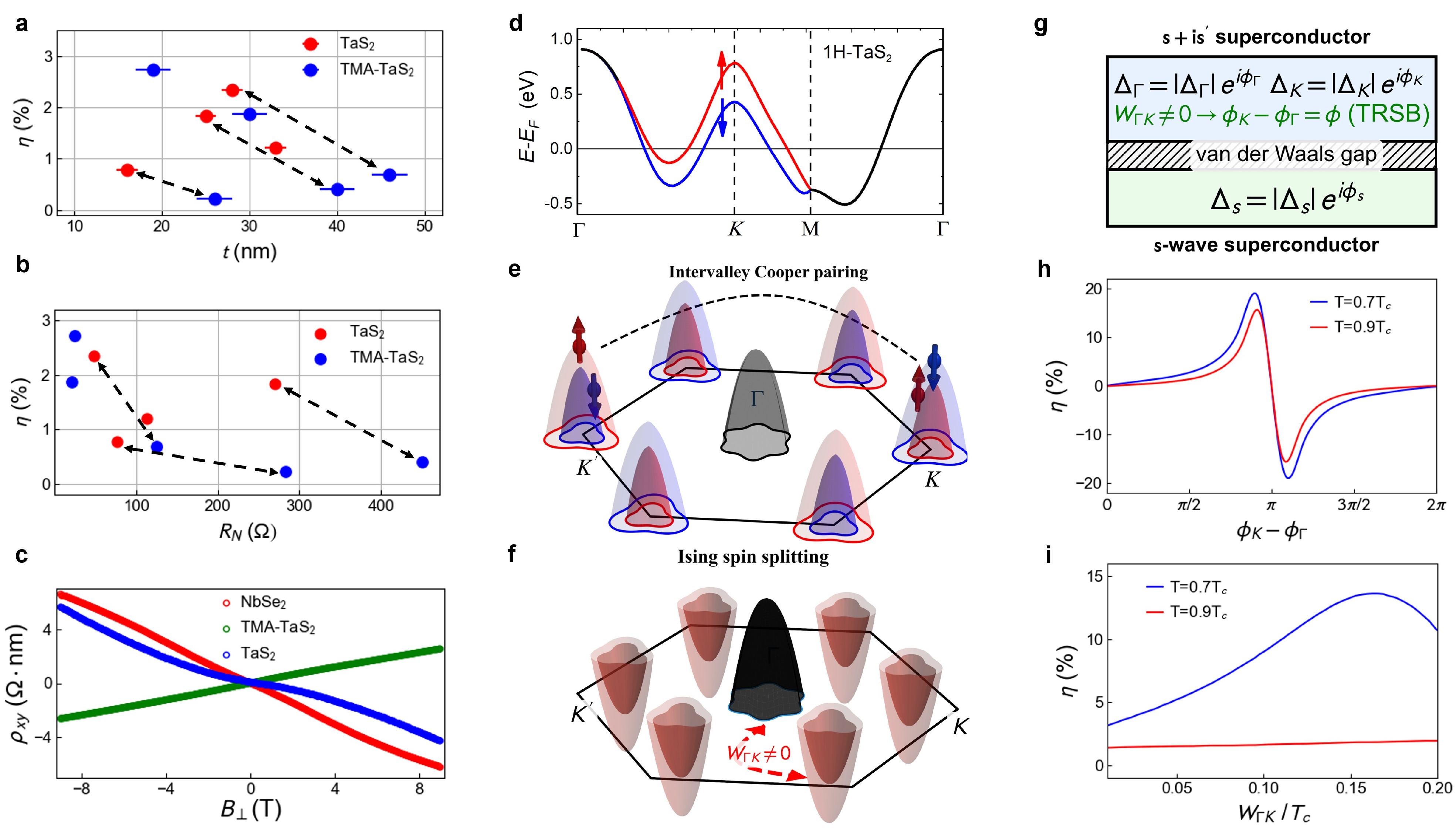}
\caption{
\textbf{Origin of the field-free Josephson diode}. \textbf{a,} Josephson diode efficiency as a function of TaS$_2$ thickness $t$. \textbf{b,} $\eta$ versus normal-state resistance R$_N$ for pristine (red) and TMA-intercalated (blue) heterojunctions. The absence of any correlation with $t$ and R$_N$ for pristine TaS$_2$ and the systematic suppression of $\eta$ upon intercalation (dashed lines link connect the same devices), rules out interfacial, strain-driven, and transparency-controlled mechanisms, pointing to an intrinsic origin of the nonreciprocal response. \textbf{c,} Transverse magnetotransport at 10 K for 2H-TaS$_2$, 2H-NbSe$_2$ and TMA-TaS$_2$. Pristine TaS$_2$ exhibits a nonlinear Hall response, revealing coexisting electron and hole carriers, while TMA-TaS$_2$ shows monolayer-like, hole-dominated transport consistent with interlayer decoupling and suppression of CDW order. \textbf{d,} DFT band structure of 1H-TaS$_2$ along the K-${\Gamma}$-M. \textbf{e,} Schematic of Ising superconductivity in TMA-TaS$_2$, highlighting spin–orbit-induced spin splitting and intervalley pairing.  \textbf{f,} Multiband superconductivity in 2H-TaS$_2$. CDW reconstruction yields electron-like states near K/K\vtick\ coexisting with a hole-like $\Gamma$-centered condensate. Interband coupling with nonzero scattering amplitude $W_{\Gamma K}$ (red line) induces an anomalous phase difference $\phi_K-\phi_{\Gamma} \neq 0,\pi$ stabilizing a TRS-breaking state. \textbf{g,} Model Josephson junction between a TRS-breaking s+is\vtick\ superconductor (TaS$_2$) and a conventional s-wave superconductor (NbSe$_2$). \textbf{h,} Calculated Josephson diode efficiency $\eta$ for two representative temperatures as a function of anomalous interband phase difference $\phi_{K}-\phi_{\Gamma}$  and fixed interband scattering amplitude $W_{\Gamma K}=0.02\,T_c$. \textbf{i,} $\eta$ versus $W_{\Gamma K}$ at fixed anomalous interband phase difference $\phi_{K}-\phi_{\Gamma}=\pi/2$.}
\label{fig4} 
\end{figure*}
We next investigate possible microscopic origins of the zero-field superconducting diode effect. Recent studies have reported spontaneous Josephson diode behavior in noncentrosymmetric van der Waals junctions containing polar weak links~\cite{Wu2022,Wu2025}. The nonreciprocal response was attributed to charge depletion regions at the interface between p-type and n-type superconductors without invoking TRS breaking~\cite {Hu2007,mis21}. In our system, 2H-TaS$_2$ and 2H-NbSe$_2$ exhibit markedly different carrier densities~\cite{Mar2025}. To test whether carrier-type asymmetry could account for the observed nonreciprocity, we engineered an effective p-n interface via molecular intercalation of TaS$_2$ (more details can be found elsewhere~\cite{mar26}).
Tetramethylammonium (TMA$^+$) cations were homogeneously intercalated between TaS$_2$ layers using a galvanic technique~\cite{Tezze2026,Mar2025}, forming a periodic superlattice of insulating molecular layers and TaS$_2$ sheets (Fig.~\ref{fig3}a). The interlayer spacing increases from $d=0.6\, \text{nm}$ to $d_{int} \approx 1\, \text{nm}$ with uniformity confirmed by optical contrast (Fig.~\ref{fig3}b,c) and transport measurements. Intercalation suppresses the charge-density wave (CDW) order and electronically decouples the layers, leading to an enhanced critical temperature consistent with monolayer-like Ising superconductivity~\cite{Mar2025}. Hall measurements (Fig.~\ref{fig3}d) show that pristine NbSe$_2$ and TaS$_2$ undergo a carrier-type crossover at the CDW transition, while intercalated TaS$_2$ remains hole-dominated over the entire temperature range due to CDW suppression. As a result, the charge gradient across the junction (Fig.~\ref{fig3}e) is enhanced upon intercalation, resulting in a p-n asymmetry at the interface. However, we find that intercalation notably reduces the nonreciprocal response (Fig.~\ref{fig3}f,g), in contrast to expectations from a charge-depletion mechanism. These observations rule out carrier-type asymmetry as the origin of the zero-field nonreciprocity.

To further examine the origin of the effect, we performed a systematic analysis of the diode efficiency $\eta$ as a function of junction parameters. No correlation is found between $\eta$ and the supercurrent density (Extended Data Fig.\,3). The self-induced magnetic field is estimated to be negligible for all devices $B_{\mathrm{self}} \lesssim 1\mu\mathrm{T}$. We also estimated the current required to nucleate a vortex/antivortex pair, $I^{v-\hat{v}} \gtrsim 10\mathrm{mA}$, which could generate nonreciprocity through current crowding, to be well above the experimental currents. These results rule out self-field and vortex-related mechanisms.

In addition, the diode efficiency shows no clear correlation with TaS$_2$ thickness or with the normal-state resistance $R_N$ (Fig.~\ref{fig4}a,b). The former excludes a purely interfacial or strain-induced mechanism, which would be expected to be enhanced in thinner systems, while the latter disfavors a transparency-driven origin of the nonreciprocal response. By contrast, molecular intercalation increases $R_N$ while reducing $\eta$, as highlighted by dashed arrows in Fig.~\ref{fig4}a,b. This behavior indicates that the nonreciprocal response is not governed by junction transparency alone, but instead reflects modifications of the intrinsic superconducting state in TaS$_2$.

In this context, we examine the electronic structure near the superconducting transition through transverse magnetotransport measurements in TMA-intercalated TaS$_2$, pristine 2H-TaS$_2$, and 2H-NbSe$_2$ (Fig.~\ref{fig4}c). TMA-TaS$_2$ exhibits a positive Hall response consistent with dominant hole transport, while TaS$_2$ and NbSe$_2$ show electron-dominated behavior at low temperatures. Notably, pristine TaS$_2$ displays a pronounced nonlinear Hall response, indicating the coexistence of electron- and hole-like carriers with distinct mobilities. Such multiband transport reflects a complex electronic structure shaped by competing orders and provides a natural platform for unconventional superconductivity.
The interplay between charge-density-wave order and superconductivity in TMDs and kagome systems has recently emerged as a fertile ground for anomalous transport phenomena. In particular, anomalous Hall effects observed in kagome superconductors~\cite{Yang2020} and TaS$_2$
 polytypes~\cite{Liu2024} have been associated with electronic states that break time-reversal symmetry in the absence of conventional magnetism~\cite{Mielke2022,chiral4Hb,sil24}. These observations suggest that intertwined multiband electronic states and competing orders can give rise to superconducting phases with nontrivial internal phase structure.

A microscopic picture consistent with these observations emerges from the band structure of TaS$_2$. In the monolayer limit, strong Ising spin-orbit coupling pins electron spins out of plane 
and spin-splits the bands around the $K/K'$  symmetry points, enabling intervalley singlet pairing (Fig.~\ref{fig4}d,e). In thin and molecule-intercalated TaS$_2$, weak interlayer coupling compared to the spin-orbit energy leads to spin-layer locking and robust Ising superconductivity~\cite{bar18,Mar2025}. In bulk 2H-TaS$_2$, inversion symmetry restores spin degeneracy~\cite{yang_enhanced_2018,xi16}, while CDW reconstruction modifies the band structure, favoring electron-like states along the $K(K')-\Gamma$ direction (Fig.~\ref{fig4}f). However, the nonlinear Hall response indicates that hole-type pockets also contribute to transport, consistent with a multiband superconducting state involving both electron- and hole-derived condensates.

Several theoretical works have suggested that multiband superconductors with interacting isotropic s-wave gaps can spontaneously break time-reversal symmetry through frustrated coupling between condensates~\cite{sta10,oiw19,yer24, yerin2014, yerin2015} or via interband scattering~\cite{yer24}. Motivated by this framework and the multiband electronic structure revealed by Hall transport, we consider a minimal phenomenological model  with two isotropic $s$-wave order parameters,
\[
\Delta_{\Gamma}=|\Delta_{\Gamma}|e^{i\phi_{\Gamma}}, \qquad
\Delta_{K}=|\Delta_{K}|e^{i\phi_{K}},
\]
associated with pairing on the $\Gamma$-centered hole pocket and the electron-like states near the $K/K'$ valleys. 
In the absence of interband scattering, the relative phase is locked at $\phi\equiv\phi_{K}-\phi_{\Gamma}=0$ or $\pi$, corresponding to conventional $s_{++}$ or $s_{+-}$ states depending on the sign of the interband coupling.   
By contrast, finite interband scattering acts as a phase-locking channel that drives the relative phase away from these values, stabilizing a complex $s+is'$ state that breaks time-reversal symmetry. More generally, an analogous anomalous phase structure may also emerge from frustration among multiple condensates.

To analyze the consequences for Josephson transport, we consider a junction between TaS$_2$, modeled as a multiband superconductor in a TRSB state, and NbSe$_2$, described to the leading approximation by a conventional $s$-wave order parameter $\Delta_s=|\Delta_s|e^{i\phi_s}$ (Fig.~\ref{fig4}g). Within the quasiclassical framework (see Ref.~\cite{yer24} and Supplementary Note~1), a finite zero-field diode effect does not arise from inversion-symmetry breaking alone. The key intrinsic ingredient is a nontrivial interband phase difference, $\phi\neq 0,\pi$, which distinguishes the TRSB multiband state from conventional $s_{++}$ and $s_{+-}$ states. However, a finite $\phi$ is not sufficient if the Josephson response remains purely sinusoidal. A diode effect emerges only when the anomalous phase structure, combined with unequal coupling of the junction to different condensates and finite interband scattering, produces a nonsinusoidal asymmetric current-phase relation, yielding $I_c^{+}\neq |I_c^{-}|$ already at zero magnetic field. In this framework, the interband scattering amplitude is the key microscopic parameter controlling the magnitude of the nonreciprocal response.

Figures~\ref{fig4}h and ~\ref{fig4}i show the calculated diode efficiency for two representative temperatures, obtained from the analytical current-phase relation derived in Supplementary Note~1. Figure~\ref{fig4}h displays $\eta(\phi)$ at fixed interband scattering amplitude $W_{\Gamma K}=0.02\,T_c$, while Fig.~\ref{fig4}i shows $\eta(W_{\Gamma K})$ at fixed anomalous interband phase difference $\phi_K-\phi_{\Gamma}=\pi/2$. In both cases, the nonreciprocal response is suppressed as temperature increases, in qualitative agreement with experiment. At fixed $W_{\Gamma K}$, $\eta$ exhibits a pronounced dependence on the intrinsic phase difference, whereas at fixed $\phi$ the interband scattering amplitude controls the magnitude of the diode effect and produces a broad maximum in $\eta$. 

This picture is qualitatively consistent with the intercalation data. While $\eta$ shows no correlation with $R_N$ in pristine junctions, intercalation leads to a concurrent decrease of $\eta$ and an increase of $R_N$. We therefore interpret the variation in $R_N$ not as the direct cause of the diode suppression, but as an accompanying signature of intercalation-induced modifications of the superconducting state, likely associated with a reduction of higher-harmonic contributions to the Josephson response. The observed zero-field rectification, together with the control experiments and the theoretical modeling, is consistent with the emergence of a nonsinusoidal asymmetric Josephson response.

Finally, we emphasize that the model used here provides a minimal phenomenological description of the field-free diode effect in 2H-TaS$_2$/2H-NbSe$_2$ Josephson junctions, where time-reversal symmetry breaking is attributed to the TaS$_2$ electrode. However, the microscopic nature of superconductivity in both
van der Waals materials remains under active investigation. In particular, a multigap description of NbSe$_2$~\cite{noat_quasiparticle_2015} is fully compatible with our model and may further enrich the Josephson response, providing additional frustration-driven routes to time-reversal-symmetry breaking and zero-field nonreciprocity. Extending this picture, our model provides a possible framework for the unconventional superconductivity reported in chiral-molecule-intercalated TaS$_2$~\cite{24_chiral} and the 4H$_b$-TaS$_2$ polytype~\cite{chiral4Hb}, in which interband scattering can be tuned through intercalation or structural modifications, thereby allowing access to superconducting states with nontrivial phase structure.\\ 
 
\noindent\large{\textbf{Conclusions}}\normalsize\\
We demonstrate a robust zero-field Josephson diode effect in TaS$_2$/NbSe$_2$ van der Waals Josephson junctions, 
providing strong evidence consistent with intrinsic time-reversal-symmetry breaking in TaS$_2$ multiband superconductor. Systematic measurements across multiple devices, including switching-current statistics, symmetry-controlled homojunctions, 
and molecular-intercalation experiments, strongly disfavor extrinsic, purely interfacial, strain-driven, and simple transparency-controlled mechanisms. 
The persistence of nonreciprocal transport therefore points to an intrinsic origin associated with the superconducting state, and may be viewed as an indirect phase-sensitive probe of the underlying order parameter. Motivated by the nonlinear Hall transport, 
we model the observations within a multiband framework in which anomalous interband phase locking, assisted by interband scattering, 
generates a nonsinusoidal asymmetric Josephson response. In this picture, a complex $s+is'$ state provides a minimal natural description of the data. More generally, our results establish Josephson nonreciprocity and anomalous Hall transport as powerful observables of emergent phase structure in multiband superconductors.

\vspace{0.5cm}

\noindent\large{\textbf{Methods}}\normalsize\\
\textbf{Sample fabrication.}
Heterostructure devices were fabricated using a polymer-based dry-transfer technique. 2H-NbSe$_2$ and 2H-TaS$_2$ crystals were mechanically exfoliated onto Si/SiO$_2$ substrates, and thin flakes were identified by optical contrast. The heterostructures were assembled using a deterministic pick-up method based on a polycarbonate (PC) film supported on a glass slide. The PC stamp was brought into contact with the selected flakes at elevated temperature ($\sim$ 90-110 Celsius degrees) to sequentially pick up and stack the layers, ensuring clean interfaces and minimizing contamination. The assembled heterostructures were then released onto pre-patterned Ti (2 nm)/Au (8 nm) electrodes on a Si/SiO$_2$ substrate. Finally, the PC film was dissolved in chloroform, yielding the completed van der Waals device. 

Selective galvanic intercalation was performed under ambient conditions by contacting a selected metal anode M$_0$ with low reduction potential to one of the electrode pads. The resulting galvanic cell with the van der Waals flake acting as the cathode was immersed in a non-aqueous electrolyte containing a salt of the target molecule. Spontaneous oxidation of the metal drives electron injection into the TaS$_2$ flake, promoting its reduction. To preserve charge neutrality, molecular cations are inserted into the van der Waals gap. Importantly, the intercalation process is governed by the relative reduction potentials of the anodic metal and the host material. As a result, the choice of the metal anode enables material-selective intercalation within van der Waals heterostructures. In particular, when using indium as the anode, intercalation occurs preferentially in TaS$_2$, while NbSe$_2$ remains unaffected, allowing selective modification of a single component in the heterostructure.

\noindent\textbf{Transport measurements.}
Electrical measurements were performed in
cryogen-free $^3$He-$^4$He dilution refrigerators equipped with two-stage RC and $\pi$ filters, using a standard four-probe configuration. Current-voltage characteristics were acquired by sweeping a low-noise DC current bias in both positive and negative directions, while the voltage drop across the junction was measured with a room-temperature low-noise pre-amplifier. 
Switching-current statistics and associated uncertainties were extracted from $1000$ repetitions of the current-voltage traces. Joule heating was minimized by automatically switching off the current once the device transitioned to the normal state. 
A delay between sweeps was optimized to maintain temperature stability within $50\, \text{mK}$. No measurable variation in the nonreciprocal response was observed across different cooling cycles, upon reversing the sweeping direction, or when introducing additional delay in the acquisition protocol. These observations indicate that local heating and instrumental artefacts are negligible.\\

\noindent\large{\textbf{Data availability}}\normalsize\\
\noindent The data that support the findings of this study are available
from the corresponding author upon reasonable request.

\bibliography{bibNbSe2}

\vspace{1cm}

\noindent\large{\textbf{Acknowledgments}}\normalsize\\
\noindent This work was supported by MICIU/AEI/10.13039/501100011033 (Grant CEX2020-001038-M), and by MICIU/AEI and ERDF/EU (Projects PID2024-157558NB-C22, PID2024-155708OB-I00, and PID2023-148225NB-C31). M.G. acknowledges support from MICIU/AEI and EU NextGenerationEU/PRTR (Fellowship RYC2021-031705-I). C.A.-G. acknowledges support from MICIU/AEI and ESF+ (Fellowship PRE2022-104487). A. A. M. and F. S. B. also thank financial support from the
European Union’s Horizon Europe research and innovation program under grant agreement No. 101130224 (JOSEPHINE). 
This work was also funded by MICIU/AEI and by the European Union NextGeneration EU Plan (PRTR-C17.I1), by the IKUR-Quantum Technologies Strategy under the collaboration agreement between DIPC, CFM, and CIC nanoGUNE on behalf of the Department of Education of the Basque Government.\\

\noindent \large{\textbf{Author contributions}}\normalsize\\
\noindent D.M. fabricated the samples, performed the experiments, analysed the data, and wrote the manuscript with inputs from all authors. D.C. and M.I. provided experimental support and optimized the acquisition systems of the low-temperature cryostat. C.A. optimized the intercalation method.  Y.Y, A. M., and S.B  conducted the theoretical modeling and wrote the theory section. M.G., S.B, L.H., and C.R. coordinated the project. All authors discussed the results and their implications equally at all stages.\\

\noindent\large{\textbf{Competing interests}}\normalsize\\
\noindent The authors declare no competing financial interests.\\

\noindent\large{\textbf{Additional Information}}\normalsize\\
\noindent 
\textbf{Supplementary Information} The online version contains supplementary material. \\

\noindent\textbf{Correspondence} and requests for materials should be addressed to Daniel Margineda and Marco Gobbi

\onecolumngrid
\newpage
\section{Supplementary Information of ``Signatures of time-reversal-symmetry breaking in multiband 2H-TaS$_2$ revealed by zero-field Josephson nonreciprocity"}

\setcounter{figure}{0}
\renewcommand{\figurename}{\textbf{
Extended Data Fig.}}

\begin{figure*}[ht]
\centering
\includegraphics[scale=0.26 ]{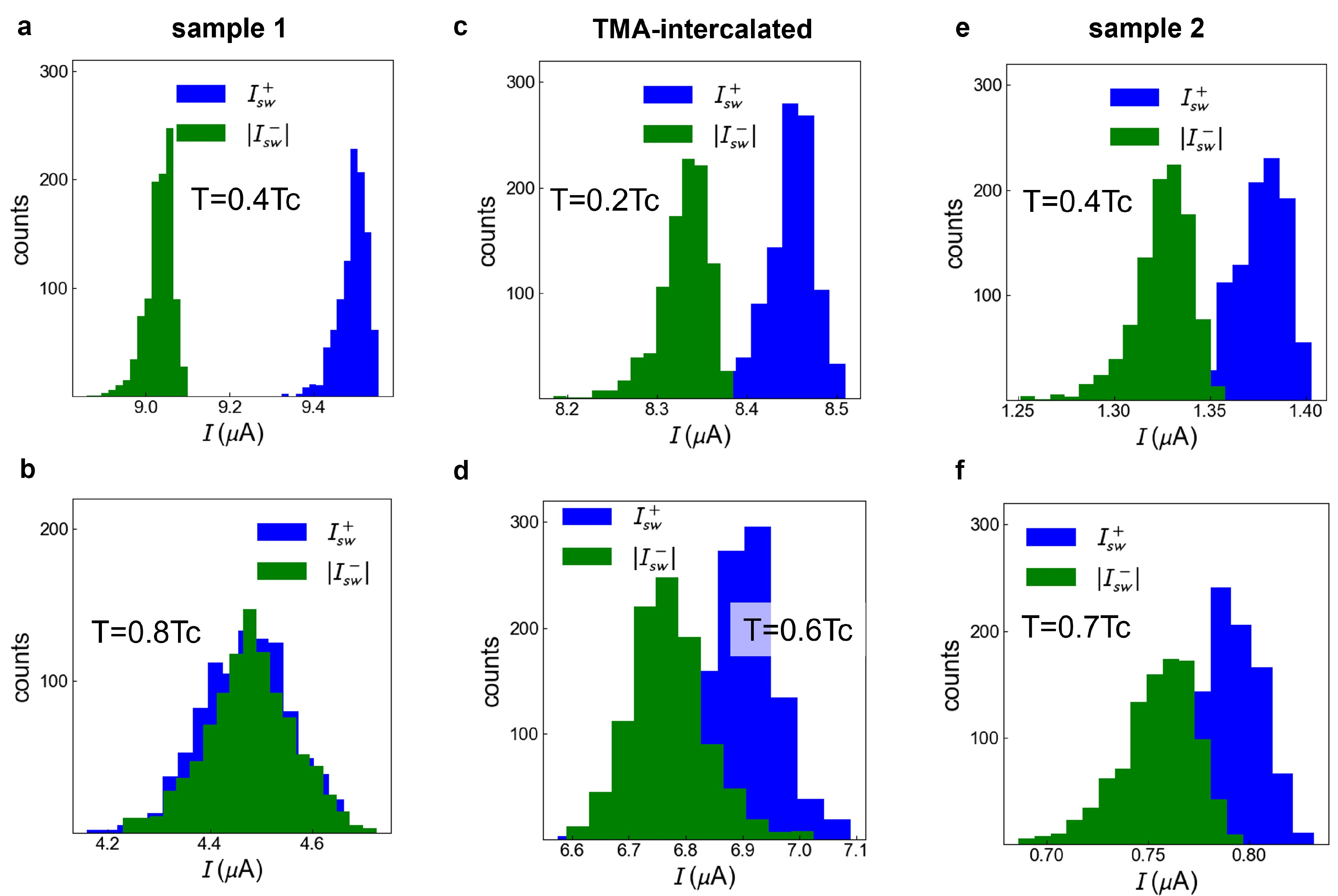}
\caption{\textbf{Temperature evolution of zero-field Josephson diode 
effect.} Positive and negative switching-current probability distribution (SCDP) of sample\,1 before \textbf{a,b} and after \textbf{c,d} selective molecular intercalation of TaS$_2$ at the indicated temperatures. Panels \textbf{e,f} show the corresponding SCPD at selected temperatures for sample\,2.}
\label{figSI1}
\end{figure*}

\begin{figure*}[ht]
\centering
\includegraphics[scale=0.26 ]{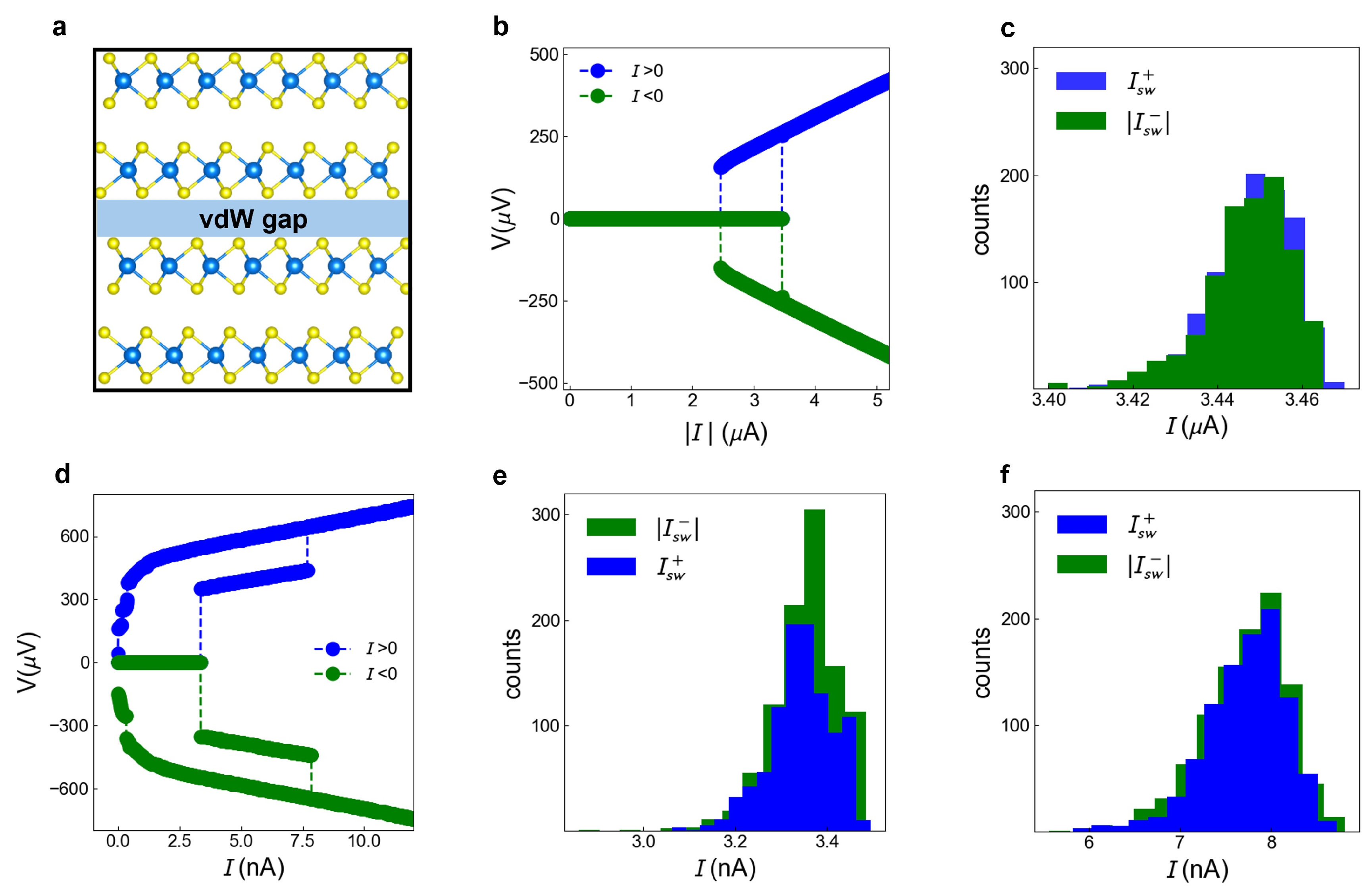}
\caption{\textbf{Reciprocal transport in TaS$_2$ Josephson homojunctions.} \textbf{a,} Schematic of the
TaS$_2$-based van der Waals Josephson junction. \textbf{b,} Zero-field current-voltage characteristic, \textbf{c,} and corresponding SCPD, measured at 100\,mK, showing reciprocal transport. \textbf{d,} Measurements in a second sample with reduced transparency, exhibiting two distinct switching currents. \textbf{e,f} SCPD corresponding to the two switching currents observed in the second sample, confirming reciprocal transport. }
\label{figSI2}
\end{figure*}
\newpage

\begin{figure*}[ht]
\centering
\includegraphics[scale=0.26 ]{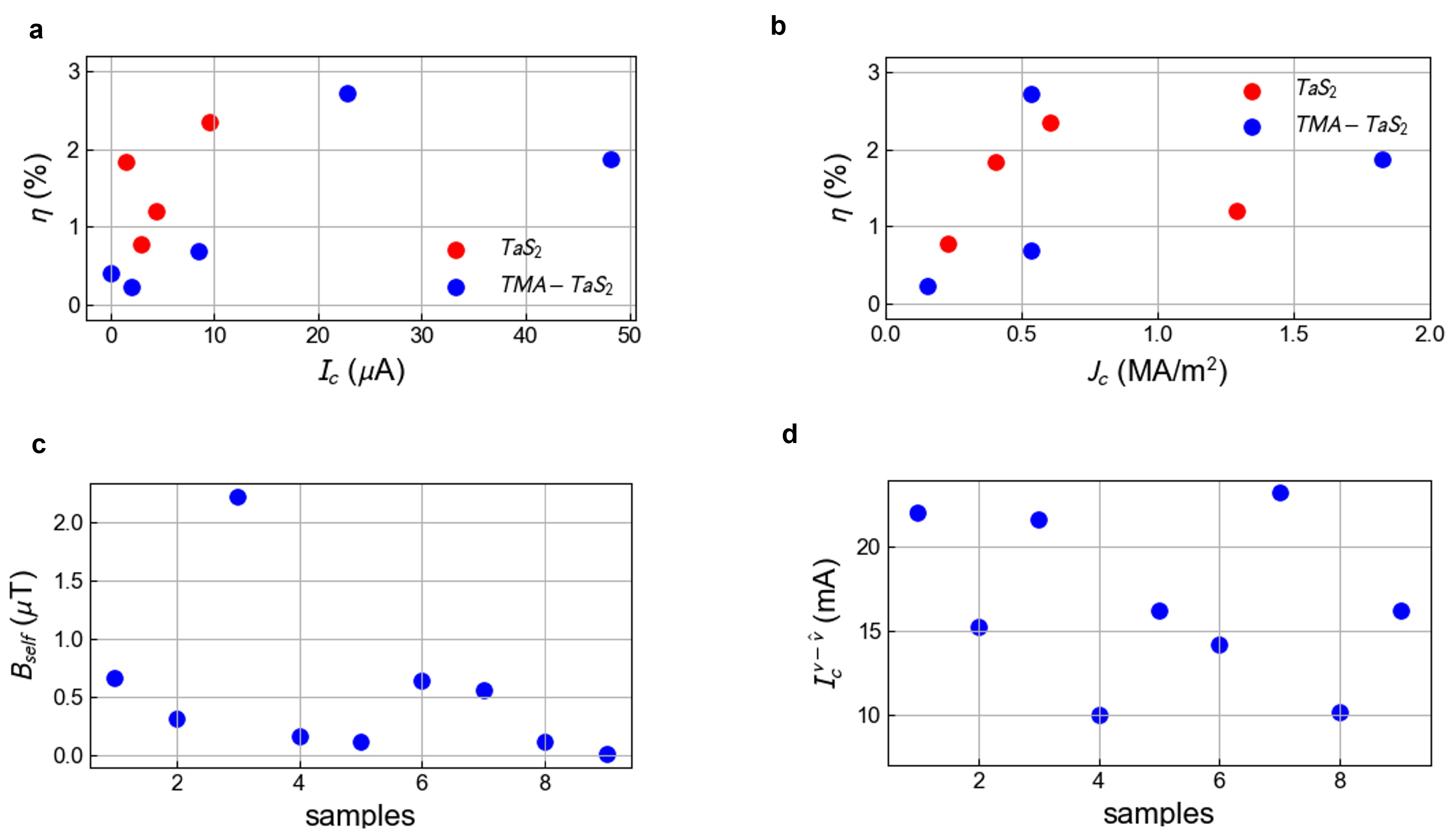}
\caption{\textbf{Ruling out extrinsic origins of the field-free Josephson diode.}  \textbf{a,} Diode efficiency as a function of the critical current. \textbf{b,} Diode efficiency as a function of the supercurrent density. The absence of a clear correlation between $\eta$ and $I_c$ excludes a twistronic origin, which would be expected to primarily control the critical current in van der Waals junctions. \textbf{c,} Estimated self-induced magnetic field produced by the critical current, well below the Earth´s magnetic field. \textbf{d,} Estimated current required to generate a vortex-antivortex pair, which could produce crowding effects, and which is well above the experimental critical currents shown in panel \textbf{a,}}
\label{figSI3}
\end{figure*}
\newpage
\clearpage
\section{Quasiclassical framework for the Josephson diode response}

We summarize the quasiclassical model underlying the theory discussion in the main text. The formulation is adapted from analytical treatment of Josephson transport between conventional and multiband superconductors, and here specialized to the minimal two-band TRSB scenario relevant for the interpretation of the TaS$_2$/NbSe$_2$ junctions.

We consider a Josephson junction between a conventional single-band $s$-wave superconductor (left bank) and a multiband superconducting terminal (right bank) in a TRSB state. In the dirty limit, the anomalous and normal quasiclassical Green functions $f_i(\mathbf{r},\omega)$ and $g_i(\mathbf{r},\omega)$ obey quasiclassical Usadel equations. In the short-junction limit, when the weak link length satisfies
\begin{equation}
L \lesssim \min_i \xi_i(T),
\end{equation}
the junction can be treated as quasi-one-dimensional and the Usadel equations reduce to
\begin{equation}
g_i \frac{d^2 f_i}{dx^2} - f_i \frac{d^2 g_i}{dx^2} = 0.
\label{eq:Usadel_short_SI}
\end{equation}

It is convenient to introduce the standard parametrization,
\begin{equation}
g_i=\frac{\omega}{\sqrt{\omega^2+\Phi_i\Phi_i^*}},
\qquad
f_i=\frac{\Phi_i}{\sqrt{\omega^2+\Phi_i\Phi_i^*}},
\qquad
\Phi_i=\frac{\omega f_i}{g_i},
\label{eq:Phi_param_SI}
\end{equation}
where $\omega = (2n+1)\pi T$ is the Matsubara frequency.

The Josephson phase difference is denoted by $\varphi$. On the left side of the junction we impose,
\begin{equation}
\Delta_0(-L/2)=|\Delta_0|e^{-i\varphi/2},
\label{eq:Delta0_bc_SI}
\end{equation}
while on the right side of the junction the two-band order parameters are written as
\begin{equation}
\Delta_1(L/2)=|\Delta_1|e^{i\varphi/2},
\label{eq:Delta1_bc_SI}
\end{equation}
\begin{equation}
\Delta_2(L/2)=|\Delta_2|e^{i\varphi/2+i\phi},
\label{eq:Delta2_bc_SI}
\end{equation}
where $\phi=\phi_{K}-\phi_{\Gamma}$ is the intrinsic anomalous interband phase difference. In the absence of interband scattering, one has $\phi=0$ or $\pi$, corresponding to conventional $s_{++}$ or $s_{\pm}$ states. For $\phi\neq 0,\pi$, the two-band state is complex and breaks time-reversal symmetry.

Close to the critical temperature, the anomalous Green functions at the right boundary can be obtained by successive expansion in the order-parameter amplitudes. To leading order one finds
\begin{equation}
f_1^{(0)}(L/2)=
\frac{(\omega+W_{21})\Delta_1+W_{12}\Delta_2}
{\omega(\omega+W_{12}+W_{21})},
\label{eq:f10_SI}
\end{equation}
\begin{equation}
f_2^{(0)}(L/2)=
\frac{(\omega+W_{12})\Delta_2+W_{21}\Delta_1}
{\omega(\omega+W_{12}+W_{21})},
\label{eq:f20_SI}
\end{equation}
where $W_{ij}$ are the interband scattering rates.

The next-order corrections read as follows,
\begin{align}
f_1^{(1)}(L/2) &=
\frac{
W_{12}(\omega+W_{21})(\Delta_1-\Delta_2)|f_2^{(0)}|^2
-
\Big[
(\omega+W_{21})^2+W_{12}(\omega+2W_{21})
\Big]\Delta_1\,|f_1^{(0)}|^2
-
W_{12}(\omega+W_{12})\Delta_2\,|f_1^{(0)}|^2
}
{\omega(\omega+W_{12}+W_{21})},
\label{eq:f11_SI}
\\[1ex]
f_2^{(1)}(L/2) &=
\frac{
W_{21}(\omega+W_{12})(\Delta_2-\Delta_1)|f_1^{(0)}|^2
-
\Big[
(\omega+W_{12})^2+W_{21}(\omega+2W_{12})
\Big]\Delta_2\,|f_2^{(0)}|^2
-
W_{21}(\omega+W_{21})\Delta_1\,|f_2^{(0)}|^2
}
{\omega(\omega+W_{12}+W_{21})}.
\label{eq:f21_SI}
\end{align}

These expressions make explicit that interband scattering already enters at the level of the boundary Green functions and provides the microscopic channel through which a nontrivial anomalous phase structure can be stabilized. For simplicity, we assume equal partial densities of states in the two bands, \(N_1=N_2\equiv N\), where \(N_1\) and \(N_2\) are the partial densities of states in bands 1 and 2, respectively. The interband scattering rates are then parameterized as \(W_{12}N_1 = W_{21}N_2 \equiv W_{\Gamma K}N\).

The Josephson current can be obtained by substituting the analytical solution of Eq.~\eqref{eq:Usadel_short_SI} into the standard expression for the quasiclassical current. In the resulting compact form, one finds,
\begin{equation}
I(\varphi)
=
\frac{\pi T}{e}
\sum_i \frac{1}{R_{N_i}}
\sum_{\omega}
\frac{C_i}{\sqrt{1-\kappa_i^2+C_i^2}}
\left[
\arctan\!\left(
\frac{\omega \kappa_i C_i + \Phi_i^- (\kappa_i^2-1)}
{\omega \sqrt{\kappa_i^2-C_i^2-1}}
\right)
-
\arctan\!\left(
\frac{\omega \kappa_i C_i + \Phi_0^- (\kappa_i^2-1)}
{\omega \sqrt{\kappa_i^2-C_i^2-1}}
\right)
\right],
\label{eq:current_total_SI}
\end{equation}
where $R_{N_i}$ are the partial normal-state resistances of the individual Josephson channels and
\begin{equation}
\kappa_i=\frac{\Phi_0^+ - \Phi_i^+}{\Phi_0^- - \Phi_i^-},
\qquad
C_i=\frac{\Phi_0^- \Phi_i^+ - \Phi_0^+ \Phi_i^-}
{\omega(\Phi_0^- - \Phi_i^-)},
\label{eq:kappaC_SI}
\end{equation}
with
\begin{equation}
\Phi_0^\pm=\frac{1}{2}\left(\Phi_0\pm \Phi_0^*\right),
\qquad
\Phi_i^\pm=\frac{1}{2}\left(\Phi_i\pm \Phi_i^*\right).
\label{eq:Phi_pm_SI}
\end{equation}

Eq. ~\eqref{eq:current_total_SI} is the analytical current-phase relation used as the basis for the theory discussion in the main text.

The key physical point is that finite inversion-symmetry breaking alone is not sufficient to produce a zero-field diode response. The essential intrinsic ingredient is a nontrivial anomalous interband phase difference $\phi\neq 0,\pi$, which distinguishes the TRSB multiband state from conventional $s_{++}$ and $s_{\pm}$ states. However, finite $\phi$ alone is also not sufficient if the current-phase relation remains purely sinusoidal. A finite diode effect emerges only when the anomalous multiband phase structure, together with unequal coupling to the different condensates and finite interband scattering, produces a nonsinusoidal asymmetric current-phase relation, yielding
\begin{equation}
I_c^+ \neq |I_c^-|.
\end{equation}

Within this framework, interband scattering acts as the key microscopic parameter controlling the onset and magnitude of the nonreciprocal response. This is the mechanism used in the main text to interpret the calculated dependences of $\eta$ on the anomalous phase difference, on interband scattering, and on temperature.

The analytical boundary functions in Eqs.~\eqref{eq:f10_SI}--\eqref{eq:f21_SI} are derived under the assumption that the order parameters of the Josephson system are small, and therefore the approach is controlled near the critical temperature of the multiband superconductor. In addition, the critical temperature itself is reduced by interband scattering. Thus, for fixed microscopic couplings, the relevant expansion is controlled in the vicinity of the scattering-renormalized critical temperature of the multiband terminal, not of the clean-limit value.

Therefore, the analytical treatment presented here should be viewed as a minimal quasiclassical framework designed to capture the qualitative dependence of the Josephson diode response on the anomalous phase difference, interband scattering, and temperature. In particular, it provides a controlled analytical route for understanding why a TRSB multiband state can naturally generate a zero-field nonsinusoidal asymmetric current-phase relation and hence a finite diode efficiency.

\end{document}